\documentclass[sigconf, nonacm]{acmart}
\usepackage{enumerate}
\usepackage[toc,page]{appendix}
\usepackage{color,graphicx,amsmath,float,subfig}
\usepackage{adjustbox}
\usepackage{cleveref}
\usepackage{enumitem}
\usepackage{lipsum}
\usepackage{multirow}
\usepackage{natbib}
\usepackage{makecell}
\usepackage{comment}

\crefformat{section}{\S#2#1#3} 
\crefformat{subsection}{\S#2#1#3}
\crefformat{subsubsection}{\S#2#1#3}

\AtBeginDocument{%
  \providecommand\BibTeX{{%
    \normalfont B\kern-0.5em{\scshape i\kern-0.25em b}\kern-0.8em\TeX}}}

\setcopyright{acmcopyright}
\copyrightyear{2018}
\acmYear{2018}
\acmDOI{XXXXXXX.XXXXXXX}

\acmConference[Conference acronym 'XX]{Make sure to enter the correct
  conference title from your rights confirmation emai}{June 03--05,
  2018}{Woodstock, NY}
\acmBooktitle{Woodstock '18: ACM Symposium on Neural Gaze Detection,
 June 03--05, 2018, Woodstock, NY} 
\acmPrice{15.00}
\acmISBN{978-1-4503-XXXX-X/18/06}

\begin{document}

\title{Think About the Stakeholders First! Towards an Algorithmic Transparency Playbook for Regulatory Compliance}

\author{Andrew Bell}
\affiliation{
  \institution{New York University}
  \streetaddress{50 West 4th St}
  \city{New York}
  \country{United States}}
\email{alb9742@nyu.edu}

\author{Oded Nov}
\affiliation{
  \institution{New York University}
  \streetaddress{50 West 4th St}
  \city{New York}
  \country{United States}}
\email{onov@nyu.edu}

\author{Julia Stoyanovich}
\affiliation{
  \institution{New York University}
  \streetaddress{50 West 4th St}
  \city{New York}
  \country{United States}}
\email{stoyanovich@nyu.edu}


\begin{abstract}

Increasingly, laws are being proposed and passed by governments around the world to regulate Artificial Intelligence (AI) systems implemented into the public and private sectors. Many of these regulations address the transparency of AI systems, and related citizen-aware issues like allowing individuals to have the right to an explanation about how an AI system makes a decision that impacts them. Yet, almost all AI governance documents to date have a significant drawback: they have focused on \emph{what} to do (or what not to do) with respect to making AI systems transparent, but have left the brunt of the work to technologists to figure out \emph{how} to build transparent systems. We fill this gap by proposing a novel stakeholder-first approach that assists technologists in designing transparent, regulatory compliant systems. We also describe a real-world case-study that illustrates how this approach can be used in practice.

\end{abstract}




\maketitle

\section{Introduction}
\label{sec:intro}

In the past decade, there has been widespread proliferation of artificial intelligence (AI) systems into the private and public sectors. These systems have been implemented in a broad range of contexts, including employment, healthcare, lending, criminal justice, and more. The rapid development and implementation of AI technologies has greatly outpaced public oversight, creating a ``wild-west''-style regulatory environment. As policy makers struggle to catch up, the issues of unregulated AI have become glaringly obvious, especially for underprivileged and marginalized communities. Famously, ProPublica revealed that the AI-driven system COMPAS used to assess the likelihood of a prisoner recidivating was highly discriminatory against black individuals~\cite{angwin2016machine}. In another example, Amazon  built and implemented an automated resume screening and hiring AI system--only to later find out that the system was biased against hiring women ~\cite{DBLP:journals/corr/abs-1909-03567}. In an effort to address these issues, countries around the world have begun regulating the use of AI systems. Over 50 nations and intergovernmental organizations have published AI strategies, actions plans, policy papers or directives ~\cite{unicri}. A survey of existing and proposed regulation around AI transparency is given in Section~\ref{sec:laws}.

Unfortunately, most strategies, directives and laws to date lack specificity on how AI regulation should be carried out \emph{in practice} by technologists. Where there is specificity, there is a lack of mechanisms for enforcing laws and holding institutions using AI accountable.  Documents on AI governance have focused on \emph{what} to do (or what not to do) with respect to AI, but leave the brunt of the work to practitioners to figure out \emph{how} things should be done~\cite{DBLP:journals/corr/abs-1906-11668}. This tension plays out heavily in regulations governing the transparency of AI systems (called ``explainability'' by AI practitioners). The most prominent example of this is the ``right to explanation'' of data use that is included in the EU’s General Data Protection Regulation (GDPR). Despite being passed into law in 2016, the meaning and scope of the right is still being debated by legal scholars, with little of the discussion resulting in concrete benefits for citizens~\cite{DBLP:conf/fat/SelbstP18}.

While regulation can help weigh the benefits of new technology against the risks, developing  effective regulation is difficult, as is establishing effective mechanisms to comply with existing regulation. This paper aims to fill a gap in the existing literature by writing to technologists and AI practitioners about the existing AI regulatory landscape, and speaks to their role in designing complaint systems. We make a case for why AI practitioners should be leading efforts to ensure the transparency of AI systems, and to this end, we propose a novel framework for implementing regulatory-compliant explanations for stakeholders. We also consider an instantiation of our stakeholder-first approach in the context of a real-world example using work done by a national employment agency.

We make the following three contributions: (1) provide a survey of existing and proposed regulations on the transparency and explainability of AI systems; (2) propose a novel framework for a stakeholder-first approach to designing transparent AI systems; and (3) present a case-study that illustrates how this stakeholder-first approach could be used in practice.

\section{Existing and Emerging Regulations}
\label{sec:laws}

In recent years, countries around the world have increasingly been drafting strategies, action plans, and policy directives to govern the use of AI systems. To some extent, regulatory approaches vary by country and region. For example, policy strategies in the US and the EU reflect their respective strengths: free-market ideas for the former, and citizen voice for the latter~\cite{gill2020policy}. Yet, despite country-level variation, many AI policies contain similar themes and ideas. A meta-analysis of over 80 AI ethics guidelines and soft-laws found that 87\% mention transparency, and include an effort to increase the explainability of AI systems~\cite{DBLP:journals/corr/abs-1906-11668}. Unfortunately, all documents to date have one major limitation: they are filled with uncertainty on \textit{how} transparency and explainability should actually be implemented in a way that is compliant with the evolving regulatory landscape ~\cite{DBLP:journals/corr/abs-1906-11668, DBLP:journals/corr/abs-1906-11668, DBLP:journals/internet/GasserA17, loi2021towards}. This limitation has 3 main causes: (1) it is difficult to design transparency regulations that can easily be standardized across different fields of AI, such as self-driving cars, robotics, and predictive modeling ~\cite{wachter2017transparent}; (2) when it comes to transparency, there is a strong information asymmetry between technologists and policymakers, and, ultimately, the individuals who are impacted by AI systems~\cite{KUZIEMSKI2020101976}; (3) there is no normative consensus around AI transparency, and most policy debates are focused on the risks of AI rather than the opportunities~\cite{DBLP:journals/internet/GasserA17}. For the purposes of scope, we will focus on regulations in the United States and Europe. However, its important noting that there is meaningful AI regulation emerging in Latin and South America, Asia, Africa, and beyond, and summarizing those regulations is an avenue for future work. For example, in 2021, Chile presented it's first national action plan on AI policy~\footnote{\url{https://www.gob.cl/en/news/chile-presents-first-national-policy-artificial-intelligence/}}.

\subsection{United States}

In 2019 the US took two major steps in the direction of AI regulation. First, Executive Order 13859 was issued with the purpose of establishing federal principles for AI systems, and to promote AI research, economic competitiveness, and national security. Importantly, the order mandates that AI algorithms implemented for use by public bodies must be ``understandable'', ``transparent'', ``responsible'', and ``accountable.'' Second, the Algorithmic Accountability Act of 2019 was introduced to the House of Representatives, and more recently reintroduced under the name Algorithmic Accountability Act of 2022. If passed into law, the Algorithmic Accountability Act would be a landmark legalisation for AI regulation in the US. The purpose of the bill is to create transparency and prevent disparate outcomes for AI systems, and it would require companies to assess the impacts of the AI systems they use and sell. The bill describes the impact assessment in detail --- which must be submitted to an oversight committee--- and states that the assessment must address ``the transparency and explainability of [an AI system] and the degree to which a consumer may contest, correct, or appeal a decision or opt out of such system or process'', 
which speaks directly to what AI practitioners refer to as ``recourse'', or the ability of an individual to understand the outcome of an AI system and what they could do to change that outcome~\cite{wachter2017counterfactual, ustun2019actionable}.

In 2019 the OPEN Government Data Act was passed into law, requiring that federal agencies maintain and publish their information online as open data. The data also must be cataloged on Data.gov, a public data repository created by the the US government. While this law only applies to public data, it demonstrates how policy can address transparency within the whole pipeline of an AI system, from the data to the algorithm to the system outcome.

There are also some industry-specific standards for transparency that could act as a model for future cross-industry regulations. Under the Equal Credit Opportunity Act, creditors who deny loan applicants must provide a specific reason for the denial. This includes denials made by AI systems. The explanations for a denial come from a standardized list of numeric reason codes, such as: ``U4: Too many recently opened accounts with balances\footnote{\url{https://www.fico.com/en/latest-thinking/solution-sheet/us-fico-score-reason-codes}}.''

\subsection{European Union}

In 2019 the EU published a white paper titled ``Ethics Guidelines for Trustworthy AI,'' containing a legal framework that outlines ethical principles and legal obligations for EU member states to follow when deploying AI\footnote{\url{https://ec.europa.eu/digital-single-market/en/news/ethics-guidelines-trustworthy-ai}}. While the white paper is non-binding, it lays out expectations on how member-states should regulate the transparency of AI systems: ``... data, system and AI business models should be transparent. Traceability mechanisms can help achieving this. Moreover, AI systems and their decisions should be explained in a manner adapted to the stakeholder concerned. Humans need to be aware that they are interacting with an AI system, and must be informed of the system’s capabilities and limitations.''

Currently, the European Commission is reviewing the Artificial Intelligence Act\footnote{\url{https://eur-lex.europa.eu/legal-content/EN/TXT/?uri=CELEX\%3A52021PC0206}}, which would create a common legal framework for governing all types of AI used in all non-military sectors in Europe. The directive takes the position that AI systems pose a significant risk to the health, safety and fundamental rights of persons, and governs from that perspective. With respect to transparency, the directive delineates between non-high-risk and high-risk AI systems (neither of which are rigorously defined at this time). It states that for ``non-high-risk AI systems, only very limited transparency obligations are imposed, for example in terms of the provision of information to flag the use of an AI system when interacting with humans.'' Yet, for high-risk systems, ``the requirements of high quality data, documentation and traceability, transparency, human oversight, accuracy and robustness, are strictly necessary to mitigate the risks to fundamental rights and safety posed by AI and that are not covered by other existing legal frameworks.'' Notably, as in the Algorithmic Accountability Act in the United States, the document contains explicit text mentioning recourse (referred to as ``redress'') for persons affected by AI systems.

The EU has also passed Regulation (EU) 2019/1150 that sets guidelines for the transparency of rankings for online search.\footnote{\url{https://eur-lex.europa.eu/legal-content/EN/TXT/?uri=CELEX\%3A32019R1150}} In practice, this means that online stores and search engines should be required to disclose the algorithmic parameters used to rank goods and services on their site. The regulation also states that explanations about rankings should contain redress mechanisms for individuals and businesses affected by the rankings.

\subsubsection{Right to Explanation.}

The Right to Explanation is a proposed fundamental human right that would guarantee individuals access to an explanation for any AI system decision that affects them. The Right to Explanation was written into the EU's 2016 GDPR regulations, and reads as follows: ``[the data subject should have] the right ... to obtain an explanation of the decision reached.''\footnote{\url{https://www.privacy-regulation.eu/en/r71.htm}} The legal meaning and obligation of the text has been debated heavily by legal scholars, who are unsure under which circumstances it applies, what constitutes an explanation~\cite{DBLP:conf/fat/SelbstP18}, and 
how the right is applicable to different AI systems~\cite{doshi2017accountability}. The Right to Explanation is an example of how emerging AI technologies may ``reveal'' additional rights that need to be considered by lawmakers and legal experts~\cite{10.1145/3306618.3314274}.

The EU's recently proposed Artificial Intelligence Act simultaneously reinforces the idea that explanations about AI systems are a human right, while slightly rolling back the Right to Explanation by acknowledging that there are both non-high-risk and high-risk AI systems. Discussions about the Right are likely to continue, and will be a central part of debates on regulating AI transparency. In fact, some local governing bodies have already taken steps to adopt the Right to Explanation. France passed the Digital Republic Act in 2016, which gives the Right to Explanation for individuals affected by an AI system in the public sector~\cite{edwards2018enslaving}. Hungary also has a similar law~\cite{malgieri2019automated}.

\subsection{Local}

There has been significant movement on the regulation of specific forms of AI systems at local levels of government. In response to the well-documented biases of facial recognition software when identifying people of different races and ethnicities~\cite{DBLP:conf/fat/BuolamwiniG18}, Washington State signed Senate Bill 6820 into law in 2020, which prohibits the use of facial recognition software in surveillance and limits its use in criminal investigation.\footnote{\url{https://app.leg.wa.gov/billsummary?BillNumber=6280&Initiative=false&Year=2019}} Detroit has also reacted to concerns about facial recognition, and its City Council approved legislation that mandates transparency and accountability for the procurement process of video and camera surveillance contracts used in the city.\footnote{\url{https://www.detroitnews.com/story/news/local/detroit-city/2021/05/25/detroit-council-approves-ordinance-boost-transparency-surveillance-camera-contracts/7433185002/}} The New York City Council recently regulated the use of AI systems in relation to employment decisions (Local Law 144 of 2021).\footnote{\url{https://legistar.council.nyc.gov/LegislationDetail.aspx?ID=4344524&GUID=B051915D-A9AC-451E-81F8-6596032FA3F9&Options=Advanced&Search}} The bill requires that AI tools for hiring employees be subject to yearly bias audits. An additional requirement is to notify job seekers that they were screened by a tool, and to disclose to them what ``qualifications or characteristics'' were used by the tool as basis of decisions. Finally, in the Netherlands, the municipality of Rotterdam has created a Data-Driven Working program which has been critical of transparency surrounding the algorithms used for fraud detection.\footnote{\url{https://nos.nl/artikel/2376810-rekenkamer-rotterdam-risico-op-vooringenomen-uitkomsten-door-gebruik-algoritmes}}

\section{The Role of Technologists}

The continuously evolving regulatory landscape of AI, combined with the limitations of existing regulation in providing clarity on how transparency should be implemented into AI systems, has left open questions concerning responsibilities for AI design and implementation. We argue that (1) practitioners should bear the bulk of the responsibility for designing and implementing compliant, transparent AI systems (2) it is in the best interest of practitioners to bear this responsibility. Researchers have also shown that there may be risks of only partially complying with AI regulations, and that fusll compliance is the best way forward~\cite{dai2021fair}. {\bf Technologists} include AI practitioners, researchers, designers, programmers, and developers.

{\bf Practitioners have the right technical expertise.} Transparency has been a central topic of AI research for the past decade, and is motivated beyond just regulatory compliance by ideas like making systems more efficient, debugging systems, and giving decision making agency to the data subjects (i.e., those affected by AI-assisted decisions) or to the users of AI systems (i.e., those making decisions with the help of AI). New technologies in transparent AI are being created at a fast pace, and there is no indication that the rapid innovation of explainable AI will slow any time soon~\cite{DBLP:conf/nips/LundbergL17, ribeiro2016should, datta2016algorithmic, DBLP:journals/corr/abs-2004-00668, DBLP:journals/corr/abs-2004-00668}, meaning that of all the stakeholders involved in the socio-technical environment of AI systems, technologists are the most likely to be aware of available tools for creating transparent AI systems. Furthermore, there are currently no objective measures for the quality of transparency in AI systems~\cite{gunning2019xai, abdul2020cogam, yang2019study, holzinger2020measuring, lu2019good}, and so technologists are necessary to discern the difference between a ``good explanation'' and a ``bad explanation'' about a system.

{\bf Practitioners are the least-cost avoiders.} This idea is based on the principle of the least-cost avoider, which states that obligations and liabilities should be allocated entirely to the party with the lowest cost of care ~\cite{stoyanovich2016revealing}. AI practitioners are the least-cost avoiders because they are already equipped with the technical know-how for building and implementing transparency tools into AI systems, especially when compared to policymakers and the individuals affected by the outcome of the system. Notably, given the wide range of existing transparency tools, implementing the ``bare minimum'' is trivially easy for most technologists.

One argument practitioners give against building transparent systems is that they may be less accurate than highly complex, black-box systems \cite{huysmans2006using}. However, there has been a growing amount of evidence suggesting that building transparent systems actually results into little to no trade-off in the accuracy of AI systems~\cite{rudin2019stop, bell2019proactive, stiglic2015comprehensible, de2018predicting}. In other words: building transparent systems is not a Pareto-reducing constraint for practitioners.

{\bf Practitioners already bear the responsibility for implementing transparency into AI systems.} A study interviewing AI practitioners found that using AI responsibly in their work is viewed as the practitioner’s burden, not the institutions for which they work. Practitioners noted that existing structures within institutions are often antithetical to the goals of responsible AI, and that it is up to them to push for structural change within that institution ~\cite{rakova2020responsible}. Section \ref{sec:laws} shows that AI regulation is converging on requiring transparent AI systems that offer meaningful explanations to stakeholders. Therefore, it is in the best interest of practitioners to continue the bottom-up approach of building transparent AI systems in the face of looming regulations.
\section{A Stakeholder-First Approach to Designing Transparent ADS}
\label{sec:explain}

\subsection{Definitions}

Technologists and AI researchers have not agreed on a definition of transparency for AI systems. Instead, a number of terms have been used, including explainability, interpretability, intelligibility, understandability, and comprehensibility~\cite{DBLP:journals/corr/abs-2012-01805}. There is no consensus on the meaning of these terms and they are often defined differently by different authors or used interchangeably. Furthermore, transparency and its related terms cannot trivially be quantified or measured, and transparency for one stakeholder does not automatically imply the same for different stakeholders~\cite{lipton2018mythos, hind2019explaining}.

While having multiple definitions of transparency has been useful for distinguishing nuance in a research setting, it also poses a challenge for policy making. In contrast to technologists, policymakers favor definitions of transparency that are about human thought and behavior such as accountability or legibility~\cite{DBLP:conf/aies/KrafftYKHB20}. Table~\ref{terms} outlines terms related to transparency commonly used by policymakers versus those used by technologists.

{\bf Transparency}. For the purposes of this paper, we choose to use only the term ``transparency,'' in the broadest possible sense, so that it encompasses all the definitions above. This is most similar to the way ``explainability'' is used by technologists.  Here we use the definition adapted from work by Christoph Molnar and define transparency as ``the degree to which a human can understand an AI system~\cite{molnar2019}.'' 

{\bf Explanation}. We use the term ``explanation'' to refer to an instantiation of transparency. For example, to ensure transparency for a system, a technologist may create an \emph{explanation} about the data it uses.

\begin{table}[h!]
\centering
\label{tab:data}
\small
\begin{tabular}{ c  c }
\toprule
{\bf Terms used by policymakers} & {\bf Terms used by technologists} \\
\midrule
\makecell{Transparency \\ Accountability \\ Understandable \\ Legibility \\ Traceability \\ Redress} & \makecell{Explainability \\ Transparency \\ Interpretablity \\ Intellegibility \\ Understandability \\ Comprehensibility \\ Recourse} \\
\bottomrule
\end{tabular}
\caption{Discrepancies in the way policymakers and AI practitioners communicate about the transparency of AI systems.}
\end{table}\label{terms}

{\bf Automated Decision Systems.} The approach described in this paper applies to all Automated Decision Systems (ADS), which is any system that processes data to make decisions about people. This means that AI systems are a subset of ADS, but there are two key distinctions: (1) an ADS is underpinned by any algorithm and not just AI or machine learning, and (2) an ADS implies a context of use and some kind of impact. For a formal definition of ADS, see ~\cite{DBLP:journals/pvldb/StoyanovichHJ20}. Henceforth, we will use the term ADS.

Notably, while many regulations are written to specifically mention ``AI systems'', all the ideas they contain about transparency could be applied to all ADS. It is likely that future regulations will focus broadly on ADS, as seen in NYC Local Law 144 of 2021 and France's Digital Republic Act.

\subsection{Running Example: Predicting Unemployment in Portugal}

To make the discussion concrete, we use a running example of an ADS implemented in Portugal to try and prevent long-term unemployment (being unemployed for 12 months or more). The long-term unemployed are particularly vulnerable persons, and tend to earn less once they find new jobs, have poorer health and have children with worse academic performance as compared to those who had continuous employment~\cite{nichols2013consequences}. The Portuguese national employment agency, the Institute for Employment and Vocational Training (IEFP), uses an ADS to allocate unemployment resources to at-risk unemployed persons. The system is based on demographic data about the individual, including their age, unemployment length, and profession, along with other data on macroeconomic trends in Portugal.

The ADS is used by job counselors who work at the IEFP unemployment centers spread across Portugal. This interaction model, where an ML system makes a prediction and a human ultimately makes a final determination informed by the system's predictions, is referred to as having a ``human-in-the-loop'' (HITL). Having a HITL is an increasingly common practice for implementing ADS ~\cite{gillingham2019can,wagner2019liable,raso2017displacement}. The ADS assigns unemployed persons as low, medium, or high risk for remaining unemployed, and then job counselors have the responsibility of assigning them to interventions such as re-skilling, resume building, or job search training ~\cite{zejnilovic2020algorithmic}.


This is a useful case study for three reasons: (1) people's access to economic opportunity is at stake, and as a result, systems for predicting long-term unemployment are used widely around the world ~\cite{platform2018tackling, sztandar2018changing, loxha2014profiling, matty2013predicting, riipinen2011risk, caswell2010unemployed}; (2) the ADS exists in a dynamic setting which includes several stakeholders, like unemployed persons, job counselors who act as the human-in-the-loop, policymakers who oversee the implementation of the tool, and the technologists who developed the tool; (3) lessons from this case about designing stakeholder-first transparent systems generalize well to other real-world uses cases of ADS.

\subsection{The Approach}
\label{sec:approach}

There are many purposes, goals, use-cases and methods for the transparency of ADS, which have been categorized in a number of taxonomies and frameworks ~\cite{DBLP:journals/jmlr/AryaBCDHHHLLMMP20, ventocilla2018towards, DBLP:journals/corr/abs-2012-01805, molnar2019, meske, DBLP:conf/chi/LiaoGM20, rodolfa2020machine, richards2021human, DBLP:journals/corr/abs-2001-09734}. The  approach we propose here has three subtle --- yet important --- differences from much of the existing work in this area: (1) our approach is \emph{stakeholder-first}, capturing an emerging trend among researchers in this field to reject existing method or use-case driven approaches; (2) our approach is focused on \emph{improving the design} of transparent ADS, rather than attempting to categorize the entire field of transparency; (3) our approach is aimed at designing ADS that comply with \emph{transparency regulations}.

\begin{figure}[t!]
    \begin{center}
    \includegraphics[width=1.0\linewidth]{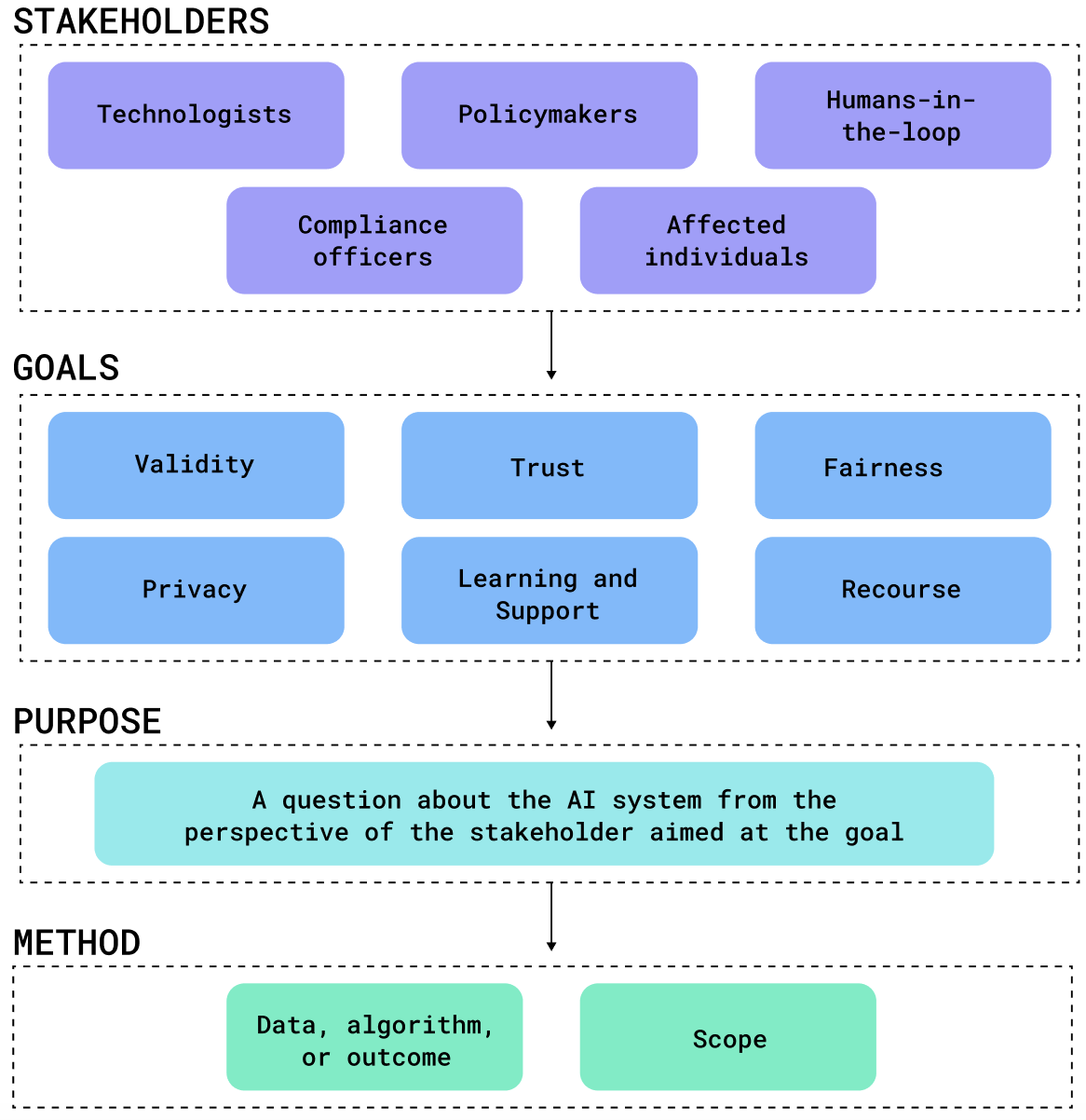}
    \end{center}
    \caption{A stakeholder-first approach for creating transparent ADS. The framework is made up of four components: stakeholders, goals, purpose, and methods. We recommend that transparency be thought of first by stakeholders, second by goals, before thirdly defining the purpose, and lastly choosing an appropriate method to serve said purpose. Using the framework is simple: starting at the top, one should consider each bubble in a component before moving onto the next component.}
    \label{fig:taxonomy}
\end{figure}

Our approach can be seen in Figure~\ref{fig:taxonomy} and is made up of the following \emph{components}: stakeholders, goals, purpose, and methods. We describe each component in the remainder of this section, and explain how they apply to the running example.

\begin{table*}[t]
\centering 
\begin{tabular}{p{0.12\textwidth}p{0.39\textwidth}p{0.39\textwidth}}
\toprule
\bf{Goal} & \bf{Definition} & \bf{Example} \\ \midrule
Validity & Making sure that an ADS is constructed correctly and is reasonable;  encompasses ideas like making sure the ADS is reliable and robust ~\cite{doshi2017towards} & An practitioner may use a transparency method to debug an ADS; An auditor may gain intuition about how an ADS is making decisions through transparency \\
Trust & Knowing ``how often an ADS is right'' and ``for which examples it is right'' ~\cite{lipton2018mythos}; influences the adoption of an ADS ~\cite{rodolfa2020machine} & A policymaker may use transparency to gain trust in the ADS; an affected individual may find through transparency that they \emph{do not} trust a particular ADS~\cite{schmidt2020transparency} \\
Fairness & Ensuring that an ADS is fair & An auditor may use an explanation about an ADS to make sure it is fair to all groups of individuals; a practitioner may use transparency tools to find bias in their modeling pipeline \\
Privacy & Ensuring that an ADS respects the data privacy of an individual & An auditor individual may use an explanation of the data used in an ADS to evaluate privacy concerns \\ 
Learning and Support &  To satisfy human curiosity, or increase understanding about how an ADS is supporting a real-world recommendation~\cite{rodolfa2020machine, molnar2019} & A doctor may use an explanation to understand an ADS recommendation of a certain treatment \\
Recourse & Allowing a stakeholder to take some action against the outcome of an ADS ~\cite{bhatt2020explainable, rodolfa2020machine} & An individual may use an explanation to appeal a loan rejection; An individual may request to see an explanation of an ADS output to understand why it was made \\
\bottomrule
\end{tabular}
\caption{Definitions and examples of stakeholder goals for the 6 categories of ADS transparency goals.}
\label{tab:goals}
\end{table*}

\subsubsection{Stakeholders} Much of ADS transparency research is focused on creating novel and innovative transparency methods for algorithms, and then later trying to understand how these methods can be used to meet stakeholders needs~\cite{bhatt2020explainable, preece2018stakeholders}. Counter to this rationale, we propose a starting point that focuses on ADS stakeholders: assuming algorithmic transparency is intended to improve the understanding of a human stakeholder, technologists designing transparent ADS must first consider the stakeholders of the system, before thinking about the system's goals or the technical methods for creating transparency.

The existing literature and taxonomies on ADS transparency have identified a number of important stakeholders, which include technologists, policymakers, auditors, regulators, humans-in-the-loop, and those individuals affected by the output of the ADS~\cite{DBLP:journals/corr/abs-2010-14374, meske, meyers2007street}.  While there is some overlap in how these stakeholders may think about transparency, in general, there is no single approach to designing transparent systems for these disparate stakeholder groups, and each of them has their own goals and purposes for wanting to understand an ADS~\cite{DBLP:journals/corr/abs-2001-09734}.  In fact, even within a stakeholder group there may be variations on how they define meaningful transparency~\cite{DBLP:conf/chi/HohmanHCDD19}. Designers of ADS may also want to weight the needs of separate stakeholders differently. For example, it may be more meaningful to meet the transparency needs of affected individuals over AI managers or auditors.

Importantly, by staking transparency on the needs of stakeholders, technologists will be compliant with citizen-aware regulations like the Right to Explanation, and those that require audits of ADS.

\emph{Running example.} In the ADS used by IEFP in Portugal, there are four main stakeholders: the technologists who developed the ADS, the policymakers who reviewed the ADS and passed laws for its implementation, the job counselors who use the system, and the affected individuals who are assessed for long-term unemployment.  In the development of the AI, explanations were created to meet the varying goals of many of these stakeholders including practitioners, policymakers, and the job counselors.  Unfortunately, and significantly, affected individuals were not considered.  Had the practitioners adopted a robust stakeholder-first approach to designing transparent systems they could have better considered how to meet the goals of this key stakeholder group. For example, a person may want to appeal being predicted low risk because they feel they are high risk for long-term unemployment and need access to better interventions.


\subsubsection{Goals.} There has been little consensus in the literature on how ADS goals should be classified. Some researchers have focused broadly, classifying the goals of ADS as evaluating, justifying, managing, improving, or learning about the outcome of an ADS ~\cite{meske}.  Others have defined goals more closely to what can be accomplished by known transparency methods, including building trust, establishing causality, and achieving reliability, fairness, and privacy ~\cite{DBLP:journals/corr/abs-2012-01805}.  Amarasinghe et al. identified five main goals (designated as use-cases) of transparency specifically in a policy setting:  model-debugging, trust and adoption, whether or not to intervene, improving intervention assignments, and for recourse. In this context, the term intervention refers to a policy action associated with the outcome of an ADS.

Notably, the goals of transparency are distinct from the purpose. The purpose addresses a context-specific aim of the ADS. For example, if an explanation is created for an ADS with the purpose of explaining to an individual why their loan was rejected, the goal may be to offer individual recourse against the rejection. This distinction is made clear in~\ref{purpose}.

For our stakeholder-first approach we make two changes to the existing body of research work. First, we require that the goal of transparent design must start with a stakeholder. Since all transparency elements of an ADS are intended for a human audience, defining a stakeholder is implicit in defining goals. Second, we have established 6 goal categories, which encompass those found in literature.  These categories are validity, trust, learning and support, recourse, fairness and privacy, and are defined in Table~\ref{tab:goals} alongside concrete examples of how these goals may be implemented.

An important discussion surrounding goals are the justifications for pursuing them.  For example, fairness and privacy goals may be justified for humanitarian reasons (they are perceived by the stakeholders as the ``right thing to do'').  Other justifications may be to prevent harm, like offering recourse to stakeholders against an outcome of an ADS, or for a reward, like an explanation that supports a doctor's correct diagnosis.  For reasons of scope we will not delve into the issue of goal justification in this paper.

\emph{Running example.} In our case study, transparency is built into the ADS with the goal of offering learning and support to job counselors. The ADS generates explanations about what factors contribute to an individual being classified as low, medium, or high risk for long-term unemployment, which job counselors use to help make better treatment decision.  Furthermore, the job counselor may also use the explanation to offer recommendations for recourse against a high risk score.


\subsubsection{Purpose} \label{purpose}  Miller proposed that the purpose of transparency is to answer a ``why'' question ~\cite{DBLP:journals/corr/Miller17a}, and gives the following example: In the context where a system is predicting if a credit loan is accepted or rejected, one may ask, ``why was a particular loan rejected?'' Liao et al. expanded on this significantly by creating a ``question bank'' which is a mapping from a taxonomy of technical transparency methodology to different types of user questions.  Instead of just answering why questions, the works shows that transparency can be used to answer 10 categories of questions:  questions about the input, output, and performance of the system, how, why, why not, what if, how to be that, how to still be this, and others ~\cite{DBLP:conf/chi/LiaoGM20}. These questions have two important characteristics. First, they are context-specific and should address a direct transparency goal of the stakeholder. Second, and importantly for technologists, these questions can be mapped onto known methods for creating explanations, meaning that a well-defined purpose for transparency acts a bridge between the goals and methods.

Thoughtfully defining the goals and purpose of transparency in ADS is critical for technologists to be compliant with regulators. It is not sufficient to try and apply general, one-size-fits-all design like simply showing the features that were used by an ADS. For instance, both the proposed Algorithmic Accountability Act in the United States and the Artificial Intelligence Act in the European Union specifically mention that ADS should have transparency mechanisms that allow individuals to have recourse against a system outcome. Researchers have noted that feature-highlighting transparency lacks utility when there is a disconnect between the explanation and real-world actions~\cite{barocas2020hidden}. For instance, if someone is rejected for a loan and the reason for that decision is the person's age, there is no action that they can effectively take for recourse against that decision.

\emph{Running example.} In the long-term unemployment use case, there were two main purposes of transparency: to understand \emph{why} an individual was assigned to a particular risk category, and to understand \emph{what} could be done to help high risk individuals lower their chances of remaining long-term unemployed.

\subsubsection{Methods.}  Once the stakeholders, goals, and purposes for algorithmic transparency have been established, it is time for the technologist to pick the appropriate transparency method (somtimes called explainablity method). Over the past decade there has been significant work in transparent ADS research (sometimes called ``explainable AI'' research or XAI) on developing new methods for understanding opaque ADS.  There are several existing taxonomies of these methods, which show that explanations can be classified on a number of attributes like the scope (local or global), intrinsic or post-hoc, data or model, model-agnostic or model-specific, surrogate or model behavior, and static or interactive ~\cite{DBLP:journals/jmlr/AryaBCDHHHLLMMP20, molnar2019, DBLP:journals/corr/abs-2012-01805}. Furthermore, researchers have created a number of different tools to accomplish transparency in ADS~\cite{DBLP:conf/nips/LundbergL17, ribeiro2016should, datta2016algorithmic, DBLP:journals/corr/abs-2004-00668, DBLP:journals/corr/abs-2004-00668}.

In contrast to the complex classification of transparency methods by technologists, regulations have focused on two elements of ADS: (1) what aspect of the ADS pipeline is being explained (the data, algorithm, or outcome)?, and (2) what is the scope of the explanation (for one individual or the entire system)? Table \ref{tab:laws} shows how different regulations speak to different combinations of pipeline and scope. In our stakeholder first-approach to transparency, we focus on these two main attributes. We will not discuss specific methods in detail, but for the convenience of technologists we have underlined them throughout this discussion.

\begin{table*}[]
\centering
\begin{tabular}{m{0.07\textwidth}m{0.24\textwidth}m{0.24\textwidth}m{0.24\textwidth}}
\toprule
 & \bf{Data} & \bf{Algorithm} & \bf{Outcome} \\
\midrule
\multicolumn{1}{l}{\bf{Local}}  & GDPR (EU) gives individuals the right to request a copy of any of their personal data & Right to Explanation gives individuals the right to know how an algorithm made a decision about them  & Both the proposed Algorithmic Accountability Act (US) and Artificial Intelligence Act (AI) give individuals the right to recourse \\
\multicolumn{1}{l}{\bf{Global}} & OPEN Government Data Act (US) mandates the government publishes public data & EU Regulation 2019/115 requires that online stores and search engines to disclose the algorithmic parameters used to rank goods and services on their site & NYC Int 1894-2020 requires hiring algorithms be audited for biased outcomes \\
\bottomrule
\end{tabular}
\caption{How different laws regulate the aspects the ADS pipeline (the data, algorithm or outcome), and within what scope (local or global).}
\label{tab:laws}
\end{table*}

\textbf{Data, algorithm, or outcome.} Transparency methods have focused on generating explanations for three different ``points in time'' in an ADS pipeline:  the data (pre-processing), the model/algorithm (in-processing, intrinsic), or the outcome (post-processing, post-hoc) ~\cite{DBLP:journals/jmlr/AryaBCDHHHLLMMP20, ventocilla2018towards}.  Importantly, transparency is relevant for each part of the machine learning pipeline because issues likes bias can arise within each component ~\cite{yang2020fairness}.

Transparency techniques that focus on the pre-processing component of the pipeline, that is, on the data used to create an ADS, typically include descriptive statistics or data visualizations.\\  
\underline{Data visualizations} have proved useful for informing users and making complex information more accessible and digestible, and have even been found to have a powerful persuasive effect ~\cite{DBLP:journals/tvcg/PandeyMNSB14, tal2016blinded}. Therefore, it is advisable to use data visualization if it can easily  address the purpose of an explanation. However, visualizations should be deployed thoughtfully, as they have the ability to be abused and can successfully misrepresent a message through techniques like exaggeration or understatement ~\cite{pandey2015deceptive}.

Techniques for creating in-processing or post-processing explanations call into question the important consideration of using explainable versus black-box algorithms when designing AI. The machine learning community accepts two classifications of models:  those that are intrinsically transparent by their nature (sometimes called directly interpretable or white-box models), and those that are not (called black box models) ~\cite{DBLP:journals/corr/abs-2012-01805}. Interpretable models, like linear regression, decision trees, or rules-based models, have \underline{intrinsic transparency mechanisms} that offer algorithmic transparency, like the linear formula, the tree diagram, and the set of rules, respectively. There are also methods like \underline{\emph{select-regress-round}} that simplify black-box models into interpretable models that use a similar set of features~\cite{jung2017simple}.


As an important design consideration for technologists, researchers have studied the effect of the complexity of a model and how it impacts its ability to be understood by a stakeholder. A user study found that the understanding of a machine learning model is negatively correlated with it's complexity, and found decision trees to be among the model types most understood by users ~\cite{DBLP:conf/scai/AllahyariL11}. An additional, lower-level design consideration is that model complexity is not fixed to a particular model type, but rather to the way that the model is constructed. For example, a decision tree with 1,000 nodes will be understood far less well than a tree with only 3 or 5 nodes.

In contrast to in-process transparency, which is intrinsically built into a model or algorithm, post-hoc transparency aims to answer questions about a model or algorithm after is has already been created.  Some of the most popular post-hoc methods are \underline{LIME, SHAP, SAGE, and QII}~\cite{DBLP:conf/nips/LundbergL17, ribeiro2016should, datta2016algorithmic, DBLP:journals/corr/abs-2004-00668}. These methods are considered \emph{model-agnostic} because they can be used to create explanations for any model, from linear models to random forests to neural networks. Some methods create a transparent \emph{surrogate model} that mimics the behavior of a black-box model. For example, \underline{LIME} creates a linear regression to approximate an underlying black-box model~\cite{DBLP:conf/nips/LundbergL17}. More work needs to be done in this direction, but one promising study has shown that post-hoc explanations can actually improve the perceived trust in the outcome of an algorithm ~\cite{DBLP:conf/softcomp/BekriKH19}.

However, post-hoc transparency methods have been shown to have two weaknesses that technologists should be aware of: (1) in many cases, these methods are at-best \emph{approximations} of the black-box they are trying to explain ~\cite{zhang2019should}, and (2) these methods may be vulnerable to adversarial attacks and exploitation ~\cite{DBLP:conf/aies/SlackHJSL20}.  Some researchers have also called into question the utility of black-box models and post-hoc explanation methods altogether, and have cautioned against their use in real-world contexts like clinical settings ~\cite{rudin2019stop}.  

\textbf{Scope}\label{subsec:scope}. There are two levels at which a transparent explanation about an ADS can operate:  it either explains its underlying algorithm fully, called a ``global'' explanation; or it explains how the algorithm operates on one specific instance, called a ``local'' explanation. Molnar further subdivides each of these levels into two sub-levels: global explanations can either be holistic (applying to an entire algorithm, which includes all of its features, and in the case of an ensemble algorithm, all of the component algorithms) or modular, meaning they explain on part of the holistic explanation and local explanations can either be applied to a single individual, or aggregated to provide local explanations for an entire group ~\cite{molnar2019}.

The scope of an explanation is highly relevant to the stakeholder and goals of an explanation, and is related to whether the stakeholder operates at a system or individual level. Researchers found that the scope of explanation can influence whether or not an individual thinks a model is fair~\cite{DBLP:journals/corr/abs-2101-09429,DBLP:conf/chi/LiaoGM20}. Policymakers and ADS compliance officers are more apt to be concerned with system level goals, like ensuring that the ADS is fair, respects privacy, and is valid overall, while humans-in-the-loop and those individuals affected by the outcome of an ADS are likely more interested in seeing local explanations to pertain to their specific cases. Technologists should consider both.

Naturally, there is considerable overlap between stakeholders' scope needs (for example, an auditor may want to inspect a model globally and look at local cases), but generally, \textit{it is important} which scope an explanation has.  Therefore designers of ADS explanations should be thoughtful of how they select the scope of an explanation based on a stakeholder and their goals.

\emph{Running-example.} In the IEFP use case, SHAP factors were given to job counselors to show the top factors influencing the score of a candidate both positively and negatively~\cite{zejnilovic2020algorithmic}. The transparency provided by SHAP provided a local explanation about the outcome of the model. A bias audit was also conducted on the entire algorithm, and presented to policy officials within IEFP.

Overall, researchers found that the explanations improved the confidence of the decisions, but counter-intuitively, had a somewhat negative effect on the quality of those decisions~\cite{zejnilovic2020algorithmic}.

\subsection{Putting the Approach into Practice}
The stakeholder-first approach describe in Section~\ref{sec:approach} is meant to act as a guide for technologists creating regulatory-compliant ADS. Putting this approach into practice is simple: starting at the first component in Figure~\ref{fig:taxonomy} (\emph{stakeholders}), one should consider each bubble, before moving onto the next component and again considering each bubble. By the time one has finished worked their way through the figure, they should have considered all the possible \emph{stakeholders}, \emph{goals}, \emph{purposes}, and \emph{methods} of an ADS. An instantiation of the approach can be found throughout Section~\ref{sec:approach} in the running example of building an ADS that predicts the risk of long-term unemployment in Portugal.

It's important to note that our proposed stakeholder-first approach is only a high-level tool for thinking about ADS transparency through the perspective of stakeholders and their needs. Beyond this approach, there are meaningful low-level steps that can be taken by technologists when it comes to actually implement transparency into ADS. One such step is the use of \emph{participatory design}, where stakeholders are included directly in design conversations~\cite{eiband2018bringing, cech2021tackling, aizenberg2020designing, gupta2020participatory}. In one promising study researchers used participatory design to successfully create better algorithmic explanations for users in the field of communal energy accounting~\cite{cech2021tackling}.



\section{Concluding Remarks}
\label{sec:conclusion}


If there is to be a positive, ethical future for the use of AI systems, there needs to be stakeholder-driven design for creating transparency algorithms --- and who better to lead this effort than technologists. Here we proposed a stakeholder-first approach that technologists can use to guide their design of transparent AI systems that are compliant with existing and proposed AI regulations. While there is still significant research that needs to be done in understanding how the transparency of AI systems can be most useful for stakeholders, and in the policy design of AI regulation, this paper aims to be a step in the right direction.  


There are several important research steps that could be taken to extend this work. First, the stakeholder-first approach described here lays the foundation for creating a complete playbook to designing transparent systems. This playbook would be useful to a number of audiences including technologists, humans-in-the-loop, and policymakers. Second, a repository of examples and use cases of regulatory-compliant systems derived from this approach could be created, to act as a reference to technologists.

\bibliographystyle{plain}
\bibliography{main}

\end{document}